\documentclass[11pt, draftclsnofoot, onecolumn]{IEEEtran}

\usepackage{amsmath,amssymb}
\usepackage[dvips]{graphicx}
\usepackage{amsfonts}
\usepackage[mathscr]{eucal}
\usepackage{latexsym}
\usepackage{amsthm}
\usepackage{exscale}
\usepackage[mathscr]{eucal}
\usepackage{bm}
\usepackage[dvipsnames]{color}
\usepackage{cases}
\usepackage{color}
\usepackage{epsfig}
\usepackage{algorithm}
\usepackage{algorithmic}
\usepackage[verbose,nospace,sort]{cite}
\usepackage{tabularx}
\usepackage{multirow}
\usepackage{multicol}
\usepackage{balance}
\usepackage{caption}
\usepackage{subcaption}
\usepackage{setspace}

\graphicspath{{./figs/}}

\begin{document}
\title{Cellular-Connected UAV: Potentials, Challenges and Promising Technologies}
\author{Yong~Zeng,~\IEEEmembership{Member,~IEEE,} Jiangbin Lyu,~\IEEEmembership{Member,~IEEE,} and Rui~Zhang,~\IEEEmembership{Fellow,~IEEE}
\thanks{Y. Zeng and R. Zhang are with the Department of Electrical and Computer Engineering, National University of Singapore, Singapore 117583 (e-mail: \{elezeng,elezhang\}@nus.edu.sg).}
\thanks{J. Lyu is with School of Information Science and Engineering, Xiamen University, China 361005 (e-mail: ljb@xmu.edu.cn).}
}
\maketitle

\begin{abstract}
Enabling high-rate, low-latency and ultra-reliable wireless communications between unmanned aerial vehicles (UAVs) and their associated ground pilots/users is of paramount importance to realize their large-scale usage in the future. To achieve this goal, {\it cellular-connected UAV}, whereby UAVs for various applications are integrated into the cellular network as new aerial users, is a promising technology that has drawn significant attention recently. Compared to the conventional cellular communication with terrestrial users, cellular-connected UAV communication possesses substantially different characteristics that bring in new research challenges as well as opportunities. In this article, we provide an overview of this emerging technology, by firstly discussing its potential benefits,  unique communication and spectrum requirements, as well as new design considerations. We then introduce promising technologies to enable the future generation of three-dimensional (3D)  heterogeneous wireless networks with coexisting aerial and ground users. Last, we present simulation results to corroborate our discussions and highlight key directions for future research.
\end{abstract}

\section{Introduction}
The past few years have witnessed a tremendous increase in the use of unmanned aerial vehicles (UAVs) in civilian applications, such as for aerial surveillance, traffic control, photography, package  delivery, and communication platforms. In June 2016, the Federal Aviation Administration (FAA) finalized the operational rules for routine commercial use of small unmanned aircraft systems (UAS). It is anticipated that the new rules will generate more than 82 billion Dollars for the U.S. economy alone and create more than 100,000 new jobs over the next decade.\footnote{https://www.inc.com/yoram-solomon/with-one-rule-the-faa-just-created-an-82-billion-market-and-100000-new-jobs.html}
 However, before the wide usage of UAVs can be practically realized, there are still many technical challenges that remain unsolved. In particular, it is of paramount importance to ensure high-capacity, low-latency and ultra-reliable two-way wireless communications between UAVs and their associated ground entities, not only for supporting their safe operation, but also for enabling mission-specific rate-demanding payload communications. However, existing UAS mainly rely on the simple point-to-point communication over the unlicensed band (e.g., ISM 2.4 GHz), which is of low data rate, unreliable, insecure, vulnerable to interference, difficult to legitimately monitor and manage, and can only operate over very limited range. As the number of UAVs and their applications are anticipated to further grow in the  coming years, it is imperative to develop new wireless technologies to enable significantly enhanced UAV-ground communications.

\begin{figure}
\centering
\includegraphics[scale=0.5]{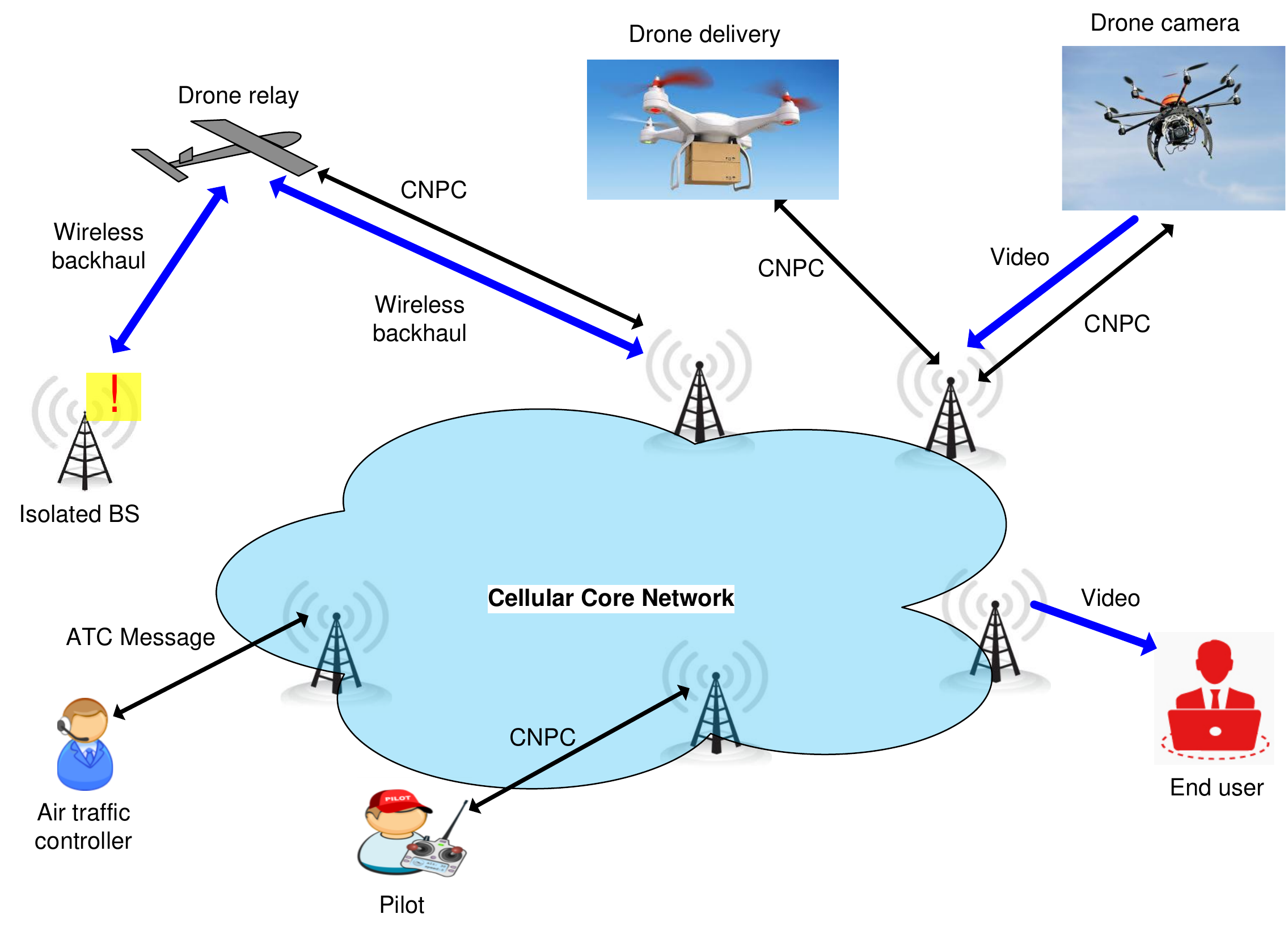}
\caption{A schematic of cellular-connected UAVs for three use cases: drone camera, drone delivery and drone wireless relaying.}\label{F:Archiecture}
\end{figure}

{\it Cellular-connected UAV} is a promising technology to achieve the above goal, whereby UAVs for various applications are integrated into the existing and future cellular networks as new aerial user equipments (UEs), as illustrated in Fig.~\ref{F:Archiecture}. Compared to the traditional  ground-to-UAV communications via point-to-point links, cellular-connected UAV has several appealing advantages, discussed as follows.

\begin{itemize}
\item {\it Ubiquitous accessibility:} {Thanks to the almost ubiquitous accessibility of cellular networks worldwide, cellular-connected UAV makes it possible for the ground pilot to remotely command and control (C\&C) the UAV with essentially unlimited operation range. Besides, it provides an effective solution to maintain wireless connectivity between UAVs and various other stakeholders, such as the end users and the air traffic controllers, regardless of their locations. For example, by leveraging cellular network, live videos can be directly sent from the UAV to distant audiences worldwide.}

\item {\it Enhanced performance:} { With the advanced cellular technologies and authentication mechanisms, cellular-connected UAV has the potential to achieve significant performance improvement over the simple direct ground-to-UAV communications,} in terms of reliability, security, and communication throughput.

\item {\it Ease of monitoring and management:} Cellular-connected UAV offers an effective means to achieve large-scale air traffic monitoring and management. {  For example, with appropriate regulations and legislations, whenever necessary, the authorized party such as the air traffic controller could legitimately take over the UAV's remote control to timely avoid any safety threat foreseen.}

\item {\it Robust navigation:} Traditional UAV navigation mainly relies on satellite such as the Global Position System (GPS), which is however vulnerable to disruption of satellite signals due to, e.g., blockage by high buildings or bad weather conditions. {  Cellular-connected UAV offers one effective method, among others such as differential GPS (D-GPS), to achieve more robust UAV navigation by utilizing cellular signals as a complementary for GPS navigation.}

\item {\it Cost-effectiveness:} {  Last but not least, cellular-connected UAV is also cost-effective. On one hand, it can reuse the millions of cellular base stations (BSs) already deployed worldwide, without the need of building new infrastructures dedicated for UAS alone, thus significantly saving the network deployment cost. On the other hand, it may also help saving the operational cost, via bundling UAV C\&C and other numerous types of payload communications into cellular systems, which will create new business opportunities for both cellular and UAV operators. Thus, cellular-connected UAV is conceived to be a win-win technology for both cellular and UAV industries, which may help facilitate the integration of UAS into the National Airspace System (NAS) cost-effectively.}
\end{itemize}

{  The attempt for supporting UAV with cellular networks can be traced back to 2000's via Global System for Mobile Communications (GSM) \cite{976}, while it has received an upsurge of interest in both academia (see, e.g., \cite{949,950,945,978,979}) and industry recently.} For instance, AT\&T and Intel demonstrated the world's first long term evolution (LTE)-connected drone at the 2016 Mobile World Congress. In August 2016, Ericsson and China Mobile conducted what they called the world's first 5G-enabled drone prototype field trial in WuXi of China. In September 2016, Qualcomm tested the drone operation over commercial LTE networks, and a trial report on LTE UAS was released in May 2017.\footnote{Qualcomm, ``LTE unmanned aircraft systems'', Trial Report, May, 2017.} In early 2017, 3GPP approved a new work item for the study on enhanced support for aerial vehicles using LTE, and a series of proposals on technical innovations have been released since then. 


Note that among the numerous UAV applications, UAV-enabled airborne communication has attracted extensive research attention recently \cite{649}, \cite{913}. Under this paradigm, dedicated UAVs are employed as aerial BSs, access points (APs), or relays, to assist the wireless communications of ground nodes, which we refer to as {\it UAV-assisted wireless communication}. The two paradigms of cellular-connected UAV communication and UAV-assisted terrestrial communication share both similarities (e.g., in terms of ground-UAV channel characteristics and interference) as well as essential differences. In particular, the roles of UAVs in these two paradigms are swapped: as BSs/APs/relays in UAV-assisted wireless communication versus  as cellular UEs in cellular-enabled UAV communication.  The main objective of this article is to give a new and state-of-the-art overview on the promising technology of cellular-connected UAV communication. The unique communication and spectrum requirements of such systems will be  discussed first, followed by the new design considerations and key promising technologies to enable our envisioned future generation of  heterogeneous wireless networks with coexisting terrestrial and aerial users in the three-dimensional (3D) space. We will further provide numerical results to corroborate our discussions and finally outline promising directions for further research.

\section{Unique Communication and Spectrum Requirement}
With UAVs as new aerial UEs, cellular-connected UAV has significantly different communication and spectrum requirements as compared to the traditional cellular communication with terrestrial users only.

\subsection{Basic Communication Requirement}
The basic communication requirements of UAS can be broadly classified into two categories: {\it control and non-payload communication} (CNPC) and {\it payload communication}.

CNPC refers to the two-way communications between unmanned aircraft and ground control station (or remote pilot) to ensure safe, reliable, and effective flight operation. Typical CNPC messages include:

\begin{itemize}
 \item Telemetry report (such as the flight altitude and velocity) from the UAV to the ground;

 \item Real-time remote C\&C for non-autonomous UAVs and regular flight command update (such as waypoint update) for (semi-)autonomous UAVs;

 \item Navigation aids as well as sense-and-avoid (S\&A) related information;

 \item Air traffic control (ATC) information relaying.
  \end{itemize}
 CNPC is usually of low data rate requirement (say, hundreds of Kbps), but has rather stringent requirement on ultra-reliability, high security, and low latency. 

On the other hand, payload communication refers to all mission-related information transmission between UAV and ground users, such as the real-time video, image, and relaying data transmission. For instance, for the particular aerial videography application, the UAV needs to timely transmit the captured video to the end users via payload communications. Compared to CNPC, UAV payload communication usually has much higher data rate requirement. For instance, to support the  transmission of full high-definition (FHD) video from the UAV to the ground user, the transmission rate is about several Mbps, while for 4K video, it is higher than 30 Mbps. The rate requirement in the UAV-enabled  airborne communication can be even higher, e.g., up to dozens of Gbps for wireless backhauling.

\subsection{Spectrum for Cellular-Connected UAV}
The lost of CNPC link has potentially catastrophic consequences. Therefore, the International Civil Aviation Organization (ICAO) has determined that CNPC links of UAV must operate over protected aviation spectrum \cite{943}. However, this was not achieved by most existing UAS with the simple direct UAV-to-ground communication over unlicensed spectrum. Furthermore, International Telecommunication Union (ITU) studies have revealed that to support CNPC for the forecasted number of UAVs in the coming years, a maximum of 34 MHz terrestrial spectrum and 56 MHz satellite spectrum is needed for supporting both LoS and beyond LoS (BLoS) UAV operations.\footnote{ITU, ``Characteristics of unmanned aircraft systems and spectrum requirements to support their safe operation in nonsegregated airspace,'' Tech. Rep. M.2171, DEC., 2009.} To meet such requirement, the C-band spectrum at 5030-5091 MHz has been made available for UAV CNPC at WRC-12 (World Radiocommunication Conference). More recently, the WRC-15 decided that ``assignments to stations of geostationary Fixed Satellite Service (FSS) networks may be used for UAS CNPC links''.\footnote{``Resolution 155 (WRC-15),'' available online at https://www.itu.int/en/ITU-R/space/snl/Documents/RES-155.pdf.}

Cellular-connected UAV has the potential to enable both LoS and BLoS UAV operations without relying on satellite. {  To meet the spectrum regulation and bandwidth requirement for CNPC, one viable approach is to license the C-band spectrum to cellular operators exclusively for CNPC links, which, together with the cellular core network, can enable BLoS C\&C for UAVs.  On the other hand, UAVs may share the common cellular spectrum pool, such as the LTE spectrum and the forthcoming 5G spectrum, with the conventional ground UEs for payload communications, as long as the interference between aerial and ground UEs is properly controlled.}

\section{New Design Considerations}\label{sec:newConsideration}
{  Cellular-connected UAV calls for a paradigm shift on the design of cellular and UAV communication systems, to enable the efficient coexistence between conventional ground UEs and the new aerial UEs. Specifically, the following new considerations need to be taken into account.}

 {\bf 3D Coverage:}
  Compared to conventional ground UEs,  UAVs typically have much higher altitude, which may even significantly exceed the BS antenna height. As a result, BSs need to be able to offer a new 3D communication coverage, as opposed to the conventional two dimensional (2D) ground coverage. However, existing BS antennas are usually tilted downwards, either mechanically or electronically, to cater for the ground coverage with reduced inter-cell interference. Despite of this, preliminary field measurement results by Qualcomm have demonstrated satisfactory aerial coverage by BS antenna sidelobes for UAVs below 120 meters (m). However, as the altitude further increases, new solutions are needed to reshape the cellular BSs to seamlessly cover the sky. For scenarios where ubiquitous 3D coverage is not attainable or simply unnecessary, e.g., for aerial pipe inspection, the concept of ``UAV highway'' may be utilized by ensuring coverage at high altitude only along certain fixed aerial corridors. {  Besides, to further improve coverage beyond sidelobes, UAV relaying via air-to-air communications will be useful.}

{\bf Unique Channel Characteristics:}
 Different from the conventional terrestrial systems, the high UAV altitude leads to unique UAV-BS channels, which usually constitute strong LoS links. For urban environment with high-rise buildings, the LoS link may be occasionally blocked, while the LoS probability typically increases with the UAV altitude. 
  Such  unique channel characteristics bring both new opportunities and challenges for the design of cellular-connected UAV communication. On one hand, the presence of LoS links usually results in strong communication channel between UAV and the associated serving BS. On the other hand, the dominance of LoS links makes the inter-cell interference a more critical issue for cellular systems with hybrid aerial and terrestrial UEs, as further discussed next.

\begin{figure}
\centering
\includegraphics[scale=0.4]{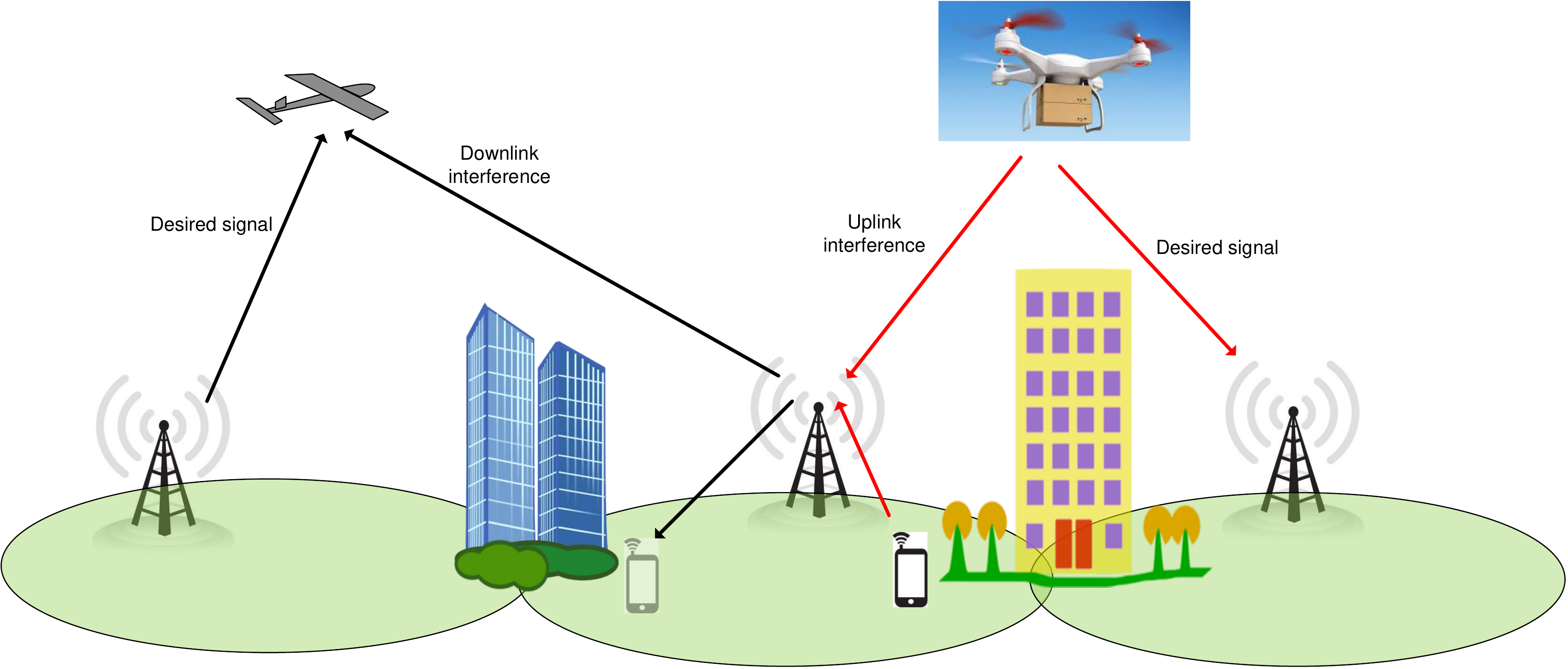}
\caption{Severe uplink UAV-to-BS and downlink BS-to-UAV interferences due to the LoS-dominated UAV-BS channels at high UAV altitude.}\label{F:AerialInterference}
\end{figure}

{\bf Severe Aerial-Ground Interference:}
  {  One major challenge to ensure the efficient coexistence between  ground  and aerial UEs lies in the severe aerial-ground interference, which is illustrated in  Fig.~\ref{F:AerialInterference}. Compared to that in conventional terrestrial systems, the interference in cellular-connected UAV systems is aggravated by the LoS-dominated UAV-BS channels at high UAV altitude.} For downlink communication from BS to UAV,\footnote{We follow the convention to use ``downlink'' to refer to the communication from BSs to UEs and ``uplink'' to that in the reverse direction, although UEs may have higher altitude than BSs in cellular-connected UAV systems.} each UAV may receive severe interference from a large number of neighboring BSs that are not associated with it, due to strong LoS-dominated channels. As a result, it is expected that an aerial UE in general would have poorer downlink performance than a ground UE, as will be verified by simulations in Section~\ref{sec:simulation}.  On the other hand, in the uplink communication from UAV to BS, the UAV could also pose strong interference to many adjacent but non-associated BSs and result in a new ``exposed BS'' interference issue. Thus, devising effective interference mitigation techniques by taking into account the unique UAV-BS channel and interference characteristics is crucial to cellular-connected UAV systems.

 {\bf Asymmetric Uplink/Downlink Traffic Requirement:}
 Different from the current cellular network, which is mainly designed for supporting the more dominant downlink (as opposed to uplink) traffic, cellular-connected UAV communications in general need to support much higher data rate in the uplink transmission from the UAV to BSs, especially for certain rate demanding applications such as video streaming and aerial imaging. {  Therefore, additional study is needed to evaluate the feasibility of supporting such asymmetric traffic requirements with existing LTE systems, with potentially dense UAV deployment. Furthermore, for future 5G-and-beyond cellular systems, new technologies can be developed to address the unique UAV traffic requirement more efficiently. One possible solution is to use drastically different bands for uplink and downlink communications, such as the conventional sub-6 GHz for UAV downlink whereas the largely under-utilized millimeter wave (mmWave) spectrum for UAV uplink.}

\section{Promising Technologies}
In this section, we discuss several promising technologies to efficiently enable future wireless systems with hybrid terrestrial and aerial users.

{\bf Sub-Sector in Elevation Domain:}
Cell sectorization is an effective technique to increase the cellular capacity by reducing inter-cell interference using directional antennas with properly designed patterns. Current cellular BSs mostly consist of three sectors along the horizontal plane, using sectorized antenna with $120^\circ$ opening. The sectorization technique can be extended to construct sub-sectors in the elevation domain for aerial users. Specifically, for each horizontal sector in current cellular systems, the intended 3D coverage volume is further partitioned into several regions (or sub-sectors) based on the elevation angles. Each sub-sector is then covered by a 3D directional antenna with appropriately designed azimuth and elevation beamwidths. 
 The sub-sector partition needs to be carefully designed by taking into account the required  coverage altitude range and the affordable UAV handoff frequency due to high mobility.

{\bf 3D Beamforming:} Compared to cell sectorization using directional antennas with fixed antenna patterns, 3D beamforming is a more flexible technique that adaptively designs the antenna beamforming based on the UAV location or even instantaneous channel state information (CSI). In this case, the BS needs to be equipped with full dimensional (FD) antenna array, such as uniform planar array (UPA), with active array elements. 
 Different from the conventional 2D beamforming with fan-shaped beam, which can be realized by 1D array such as uniform linear array (ULA), 3D beamforming offers more refined angle resolutions in both azimuth and elevation dimensions and results in pencil-shaped beam. This thus helps significantly enhance the interference mitigation capability by exploiting the elevation angle separations of UAVs. Note that 3D beamforming has also received notable interest in conventional cellular networks \cite{946}. However, the large variation range of the elevation angles of UAVs and the dominance of the LoS UAV-BS channels make 3D beamforming especially appealing for cellular-connected UAV systems. Specifically, compared to conventional cellular networks with terrestrial UEs only, it is more likely to find two UEs with sufficiently separated elevation angles in systems with both aerial and ground UEs, thus making 3D beamforming more effective. 

{\bf Multi-Cell Cooperation:} The LoS-dominating characteristic of UAV-BS channels brings new opportunities for multi-cell cooperation. Specifically, as the UAV at higher altitude is likely to have strong LoS links with more neighboring BSs, cellular-connected UAVs potentially can enjoy larger macro-diversity gain brought by multi-cell cooperation than conventional terrestrial users. There are in general two forms of multi-cell cooperation: {\it coordinated resource allocation} and {\it coordinated multi-point (CoMP) transmission/reception}. With coordinated resource allocation, the communication resources such as channel assignment, power allocation, beamforming weights and UE-BS association, are jointly optimized across different cells by taking into account the co-channel interference. {  One viable coordination technique is interference alignment, which could be particularly useful to cancel out the LoS interference for UAV systems.} On the other hand, with CoMP transmission/reception, the signals for each UE are jointly transmitted/received by multiple cooperating BSs that form a virtual distributed antenna array. 

 While multi-cell cooperation has been extensively studied in conventional cellular systems, its implementation for cellular systems with hybrid aerial and terrestrial users faces new design challenges. In particular, the set of cooperating BSs need to be carefully chosen to achieve a desired trade-off between performance and backhaul overhead, by taking into account the flying status such as UAV speed and altitude, as well as the unique UAV-BS channel characteristics. One appealing approach is to apply the ``UAV-centric'' cell cooperation, where larger-scale multi-cell cooperation is applied for those UAVs with low speed (or relatively slow channel variations) and/or at high altitude (with potentially large macro-diversity gains).

{\bf Ground-Aerial NOMA:} Non-orthogonal multiple access (NOMA) is a promising technology to increase the spectrum efficiency in 5G cellular systems \cite{947}. 
 Studies have revealed that compared to conventional orthogonal multiple access (OMA), NOMA yields the highest performance gain when the channel conditions of the users are most different \cite{947}. This makes NOMA a very attractive technology for simultaneously serving the payload communications of UAVs and ground UEs, referred to as {\it ground-aerial NOMA}, thanks to the generally asymmetric channel conditions for these two types of UEs. Consider the power-domain uplink ground-aerial NOMA as an example. As UAVs at high altitude typically have much stronger LoS communication links with the BS than ground UEs given the similar distance, the BSs could first decode the signal from the UAVs while treating that from the ground UEs as noise, and then subtract the decoded UAV signals before decoding the weaker signals for the ground UEs. 

\section{Simulation Results and Discussions}\label{sec:simulation}
\begin{figure}
\centering
\begin{subfigure}{0.45\textwidth}
\centering
\includegraphics[width=\textwidth]{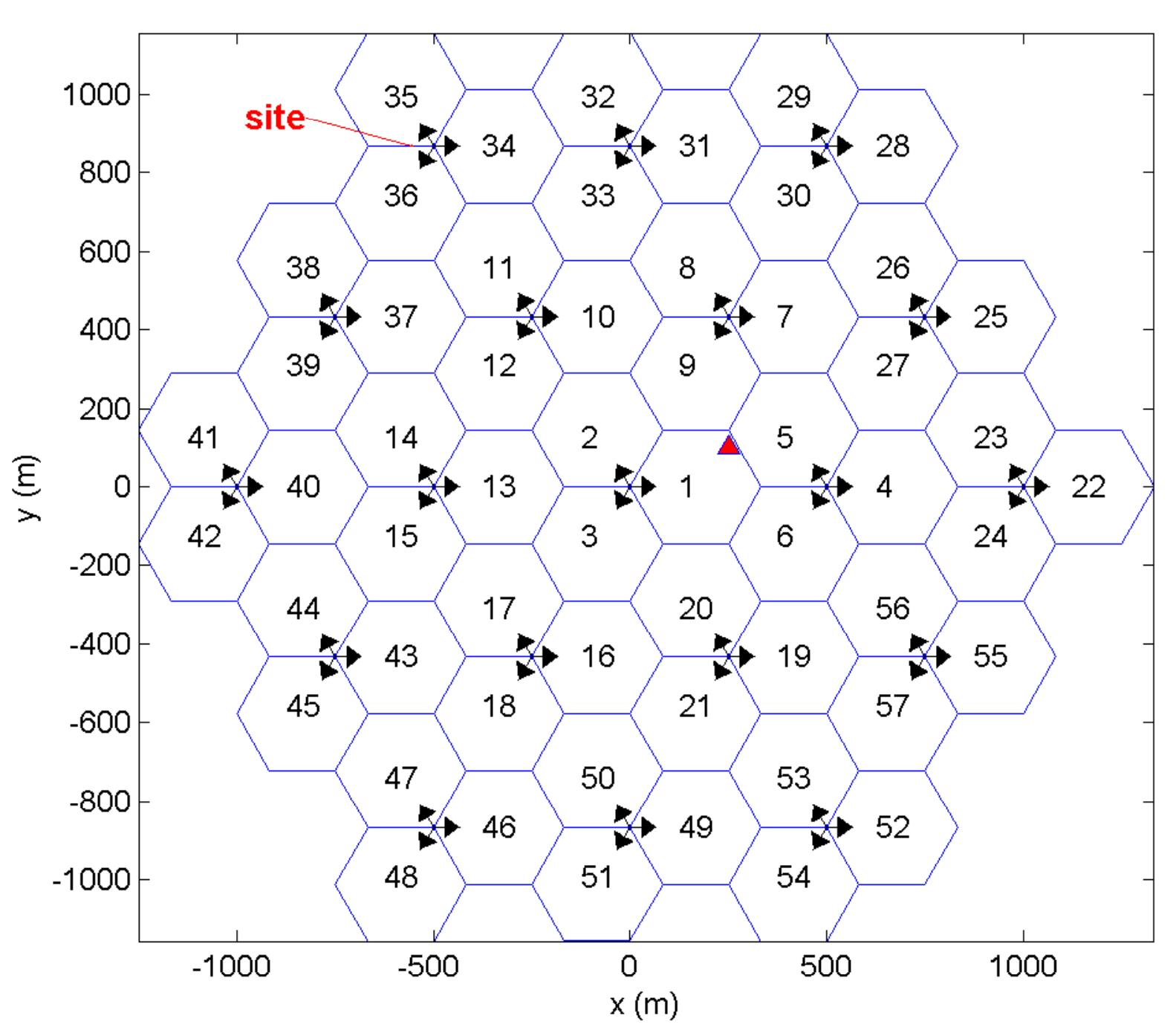}
\caption{Cell layout. Arrows denote boresight of each cell.}
\end{subfigure}
\hspace{0.02\textwidth}
\begin{subfigure}{0.5\textwidth}
\centering
{\footnotesize \setstretch{1} \begin{tabular}{p{0.32\textwidth}|p{0.68\textwidth}}
\hline
BS antenna height & 25 m \\
\hline
Carrier frequency & 5 GHz for UAV C\&C and 2 GHz for others \\
\hline
Channel bandwidth & 1 MHz \\
\hline
Transmit power by each cell& 20 dBm, equally allocated among associated UEs \\
\hline
BS antenna element pattern & 3GPP TR38.901 V14.0.0 \\
\hline
 Array configuration at each cell & {\it Fixed pattern:} $8 \times 1$ ULA, $10^\circ$ downtilt; \newline {\it 3D beamforming:} $8\times 4$ UPA\\
\hline
Channel modeling & LoS probability, pathloss and shadowing: 3GPP R1-1714856; \newline  Small-scale fading: 3GPP TR38.901 V14.0.0 with $K=15$ dB\\
\hline
Cell association & {\it Fixed pattern:} Maximum RSRP based on large-scale channel gain; \newline {\it 3D beamforming:} Maximum RSRP with MRT beamforming based on instantaneous CSI \\
\hline
Noise power spectral density & $-174$ dBm/Hz, with 9 dB noise figure \\
\hline
\end{tabular}}
\caption{System parameters.}
\end{subfigure}
\caption{Simulation setup.}\label{F:setup}
\end{figure}

 As shown in Fig.~\ref{F:setup}(a), we consider a cellular system with 19 sites, each constituting 3 sectors/cells so that a total of 57 cells are considered. The cell indices are labeled in the figure. The 3GPP urban macro (UMa) scenario is considered, where the inter-site distance (ISD) is 500 m and the corresponding cell radius is 166.7 m. We consider two different array configurations at each cell: {\it fixed pattern} versus {\it 3D beamforming}. With fixed pattern, a ULA of size  $(M,N)=(8,1)$ is employed at each sector, where $M$ and $N$ denote the number of antenna elements along the vertical and horizontal dimension, respectively. For this configuration, the steering magnitude and phase of each antenna element is fixed to achieve a $10^\circ$ electrical downtilt. On the other hand, with 3D beamforming, each sector is equipped with a UPA of size $(M,N)=(8,4)$, and the signal magnitude and phase by each antenna element can be flexibly designed to enable 3D beamforming.  For both array configurations, adjacent antenna elements are separated by half wavelength, and the antenna element pattern follows from the 3GPP technical specification\footnote{3GPP TR38.901, ``Study on channel model for frequencies from 0.5 to 100 GHz'', V14.0.0.}, with half-power beamwidth given by $65^\circ$ both along the azimuth and elevation dimensions.  All relevant system parameters are summarized in Fig.~\ref{F:setup}(b).

\subsection{UAV C\&C with Dedicated Channel}
First, we consider the downlink C\&C communication from the BS to the UAV. As marked by the red triangle in Fig.~\ref{F:setup}(a), we focus on one particular UAV with horizontal coordinate (250 m, 100 m), i.e., a UE near the edge of cells 1, 5 and 9. {  We assume that one dedicated channel at C-band with carrier frequency 5 GHz is exclusively assigned for this UAV. Therefore, interference issue is not present in this example.} Three UAV altitude values are considered: $H_{\mathrm{ue}}=$ 1.5 m, 90~m, and 200 m. Note that the UE altitude of 1.5 m may correspond to either a benchmark ground UE or a UAV in take-off/landing status.

\begin{figure}
\centering
\begin{subfigure}{0.48\textwidth}
\includegraphics[width=\linewidth]{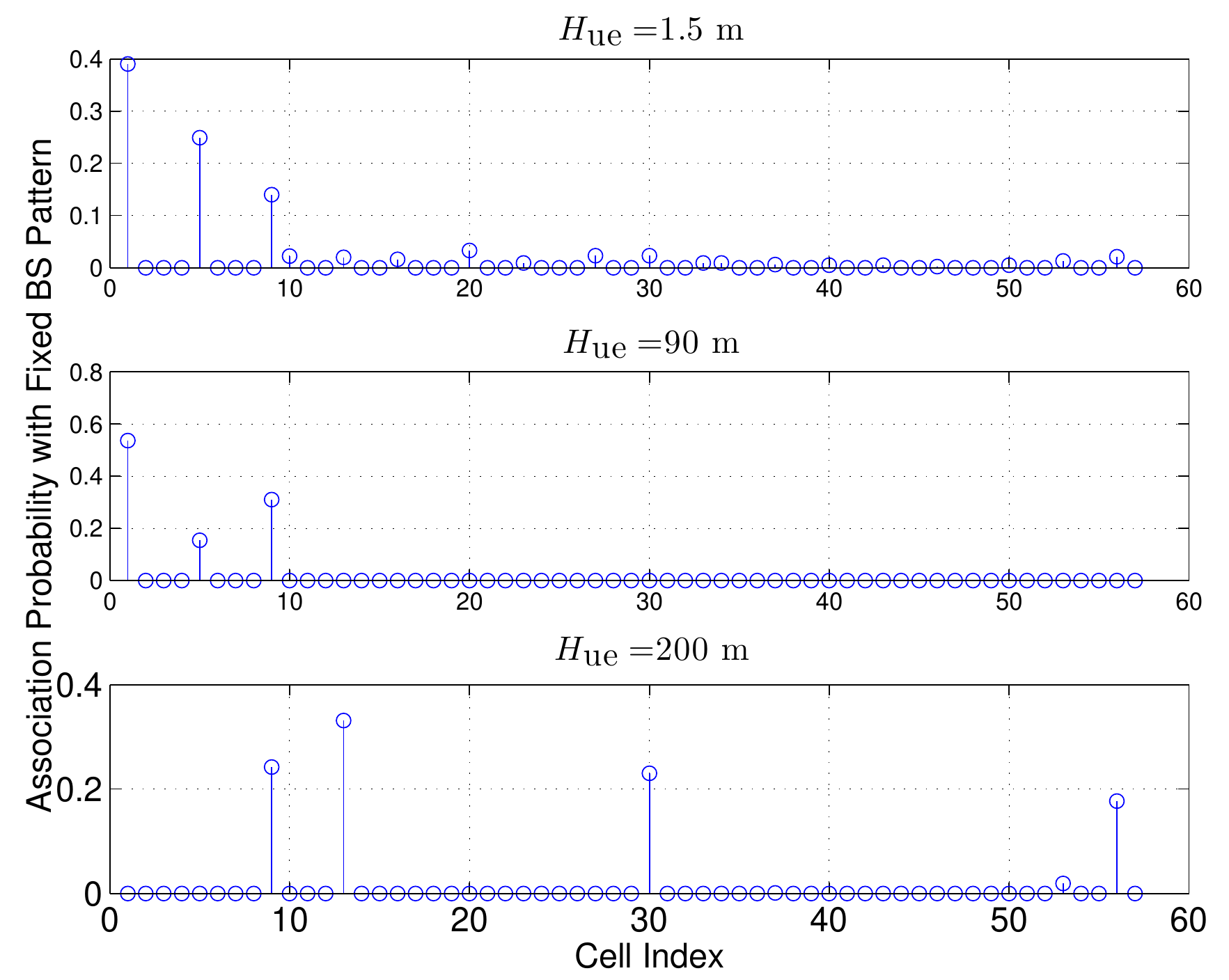}
\caption{Fixed BS pattern}
\end{subfigure}
\hspace{0.02\textwidth}
\begin{subfigure}{0.48\textwidth}
\centering
\includegraphics[width=\linewidth]{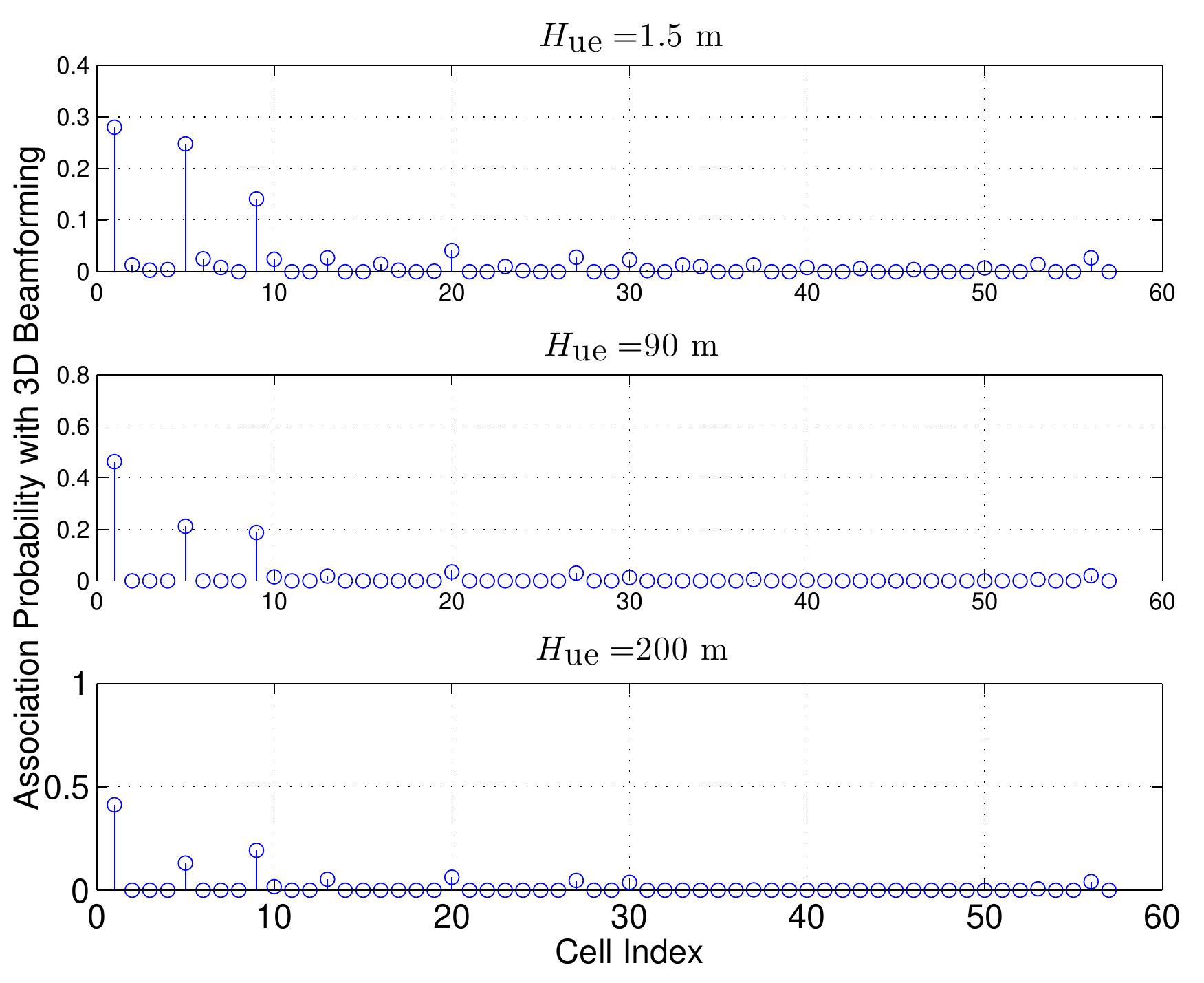}\\
\caption{3D Beamforming}
\end{subfigure}
\caption{Association probability at different UAV altitude.}\label{F:AssoProb}
\end{figure}

Fig.~\ref{F:AssoProb} shows the cell association probability at different UAV altitude, where the maximum reference signal received power (RSRP) association rule is assumed. For fixed BS pattern, the RSRP is calculated based on the large-scale channel gain (pathloss and shadowing), whereas for 3D beamforming,  it is obtained via the maximal-ratio transmission (MRT) beamforming based on instantaneous CSI. It is observed from Fig.~\ref{F:AssoProb}(a) that with fixed BS pattern, the UAV is most likely to be associated with the nearby cells when the altitude is low (e.g., cells 1, 5 and 9 for $H_{\mathrm{ue}}=$ 1.5 m and 90 m). However, as the altitude increases, it is more likely that the associated cell is far away from the UAV, e.g., cells 13, 30 and 56 for $H_{\mathrm{ue}}=$ 200 m. This is expected since for UAV at higher altitude, the elevation angle-of-departures (AoDs) with respect to those nearby BSs are larger, and hence the UAV may more easily fall into the antenna nulls of those nearby BSs due to the downtilted antenna pattern. As a result, the UAV has to be associated with those distant cells via their antenna side lobes. {  Note that similar association results exist for UAV-assisted wireless communications with UAVs acting as aerial platform to serve ground UEs \cite{977}.} In contrast, with 3D beamforming, it is observed from Fig.~\ref{F:AssoProb}(b) that the UAV is almost surely associated with the nearby cells even for high altitude at $H_{\mathrm{ue}}=$ 200 m, thanks to the flexible beam adjustment to focus signals to the UAV with 3D beamforming.

\begin{figure}
\centering
\begin{subfigure}{0.48\textwidth}
\centering
\includegraphics[width=\linewidth]{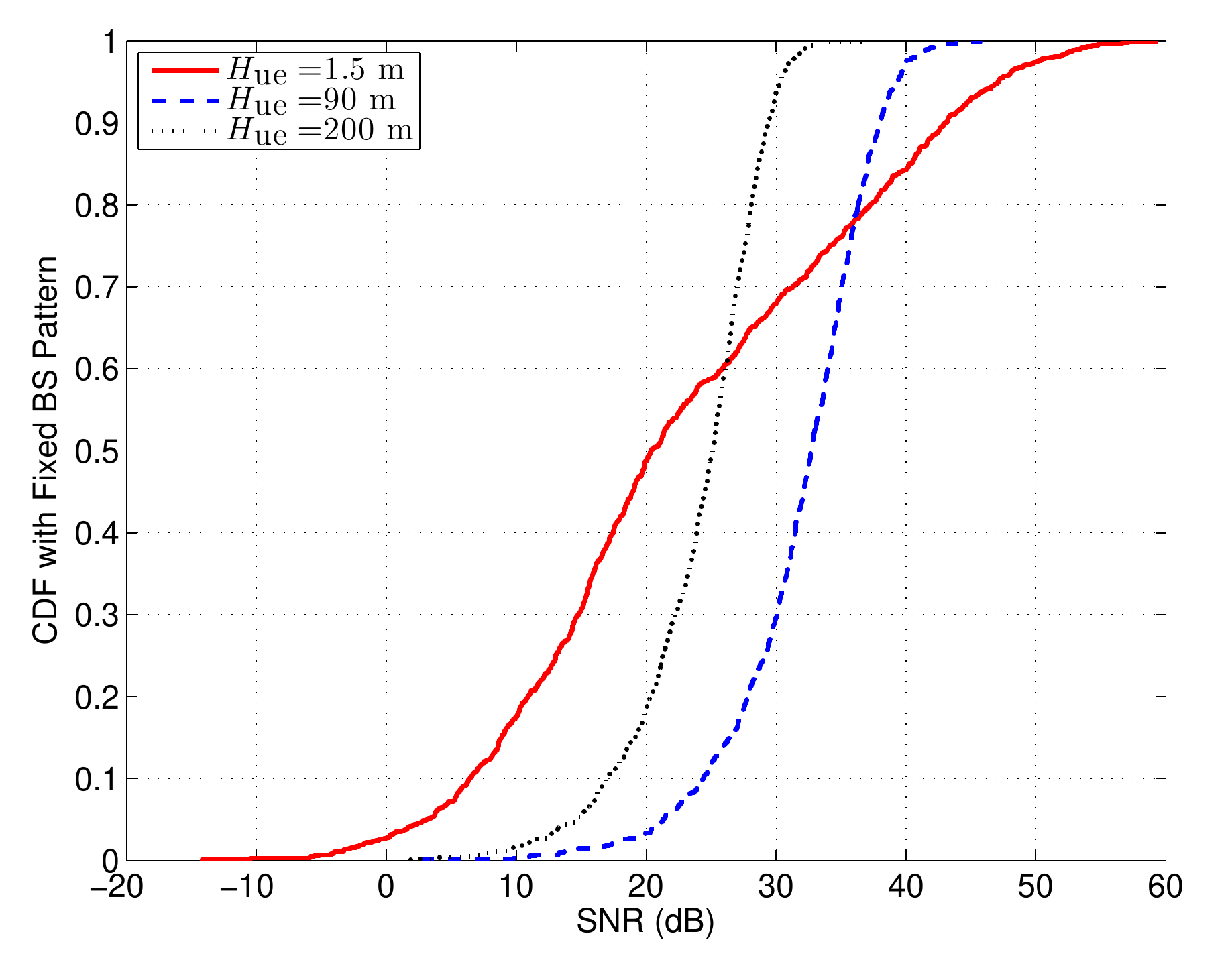}
\caption{Fixed BS pattern}
\end{subfigure}
\hspace{0.02\textwidth}
\begin{subfigure}{0.48\textwidth}
\includegraphics[width=\linewidth]{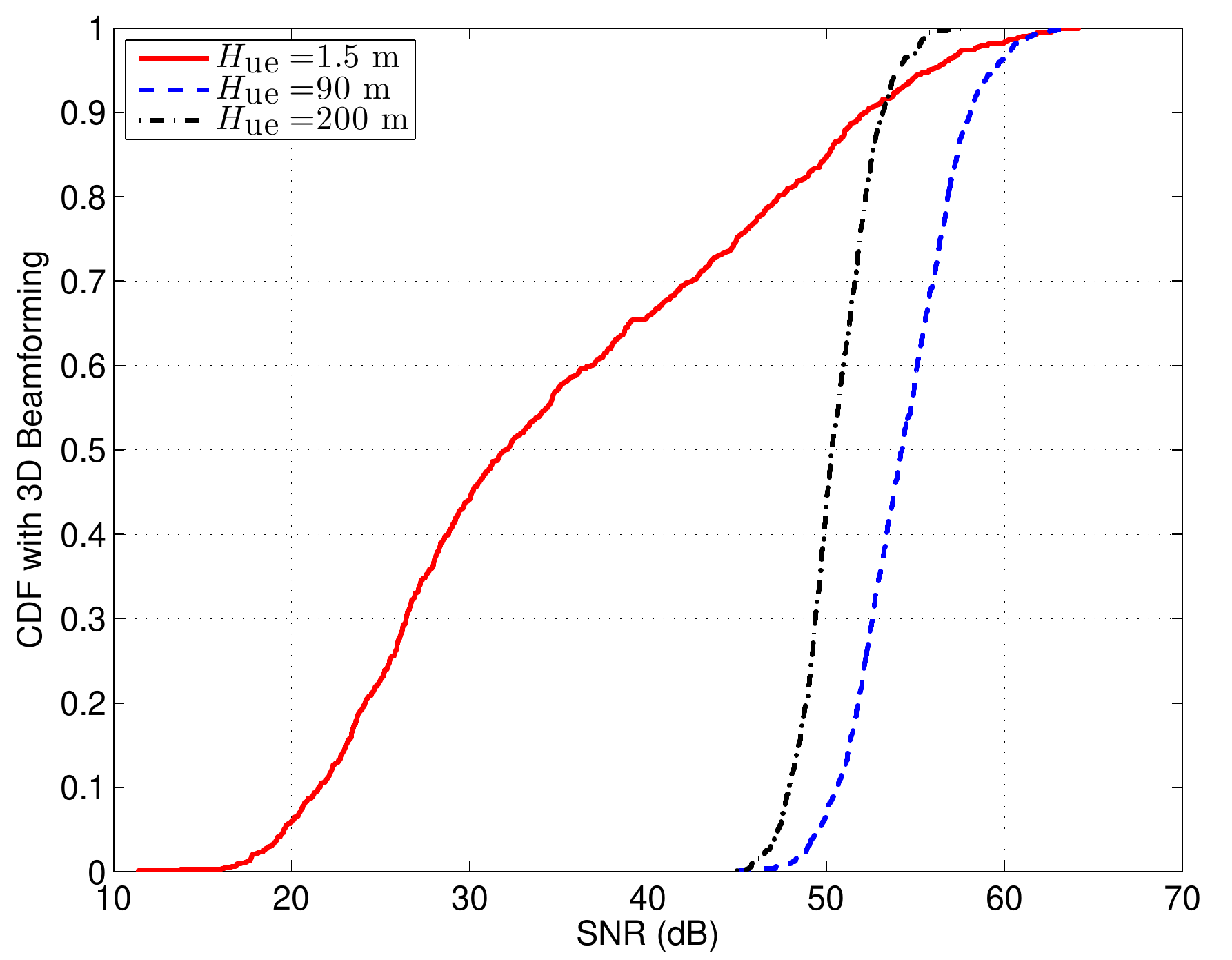}
\caption{3D beamforming}
\end{subfigure}
\caption{Empirical CDF of SNR at different UAV altitude.}\label{F:SNRCDF}
\end{figure}

Fig.~\ref{F:SNRCDF} plots the empirical cumulative distribution function (CDF) of the received signal-to-noise ratio (SNR) at the UAV. It is observed that for both fixed pattern and 3D beamforming, higher UAV altitude leads to less SNR variations, which is expected due to the higher LoS probability as the altitude increases. Besides, it is observed that compared to UE at the ground level (i.e., $H_{\mathrm{ue}}=1.5$ m), the worst-case SNRs are significantly improved at high altitude. For example, even with fixed BS pattern, the 5th percentile SNR (below which $5\%$ of the observations are found)  with $H_{\mathrm{ue}}=$ 200 m is about $10$dB higher than that at $1.5$ m, though in the former case, the UAV has to be associated with the more distant cells via their sidelobes as shown in Fig.~\ref{F:AssoProb}(a). This result demonstrates that the benefit of LoS communication link at high UAV altitude well compensates the small antenna gain of the sidelobe, thus validating the feasibility of leveraging existing cellular systems for UAV C\&C at moderate UAV altitude. Furthermore, for $H_{\mathrm{ue}}=$200 m, the SNR is lower than that for $H_{\mathrm{ue}}=90$ m, due to the increased link distance as well as the reduced side lobe gain with increasing $H_{\mathrm{ue}}$. Moreover, by comparing Fig.~\ref{F:SNRCDF}(a) and Fig.~\ref{F:SNRCDF}(b), it is observed that 3D beamforming is able to significantly improve the SNR performance for all UAV altitude, and the performance improvement is more significant for $H_{\mathrm{ue}}=200$ m due to the more dominant LoS communication link.

\subsection{Shared Channel by UAV and Ground UE}
{  Next, we consider the multi-user downlink payload communication from BSs to UAVs. We assume that the UAVs reuse the same set of channels with the conventional ground UEs.} We focus on one particular channel that is reused by a total of 20 UEs, out of which $N_{\mathrm{UAV}}$ are aerial UEs. Three values for $N_{\mathrm{UAV}}$  are considered: $N_{\mathrm{UAV}}$= 0, 5 and 10. The horizontal locations of all UEs are uniformly distributed in a circular disk with radius 1000 m. For ground UEs, the altitude is fixed to 1.5 m, whereas for aerial UEs, it is uniformly distributed between 1.5 m and 300 m.

\begin{figure}
\centering
\includegraphics[width=0.6\linewidth]{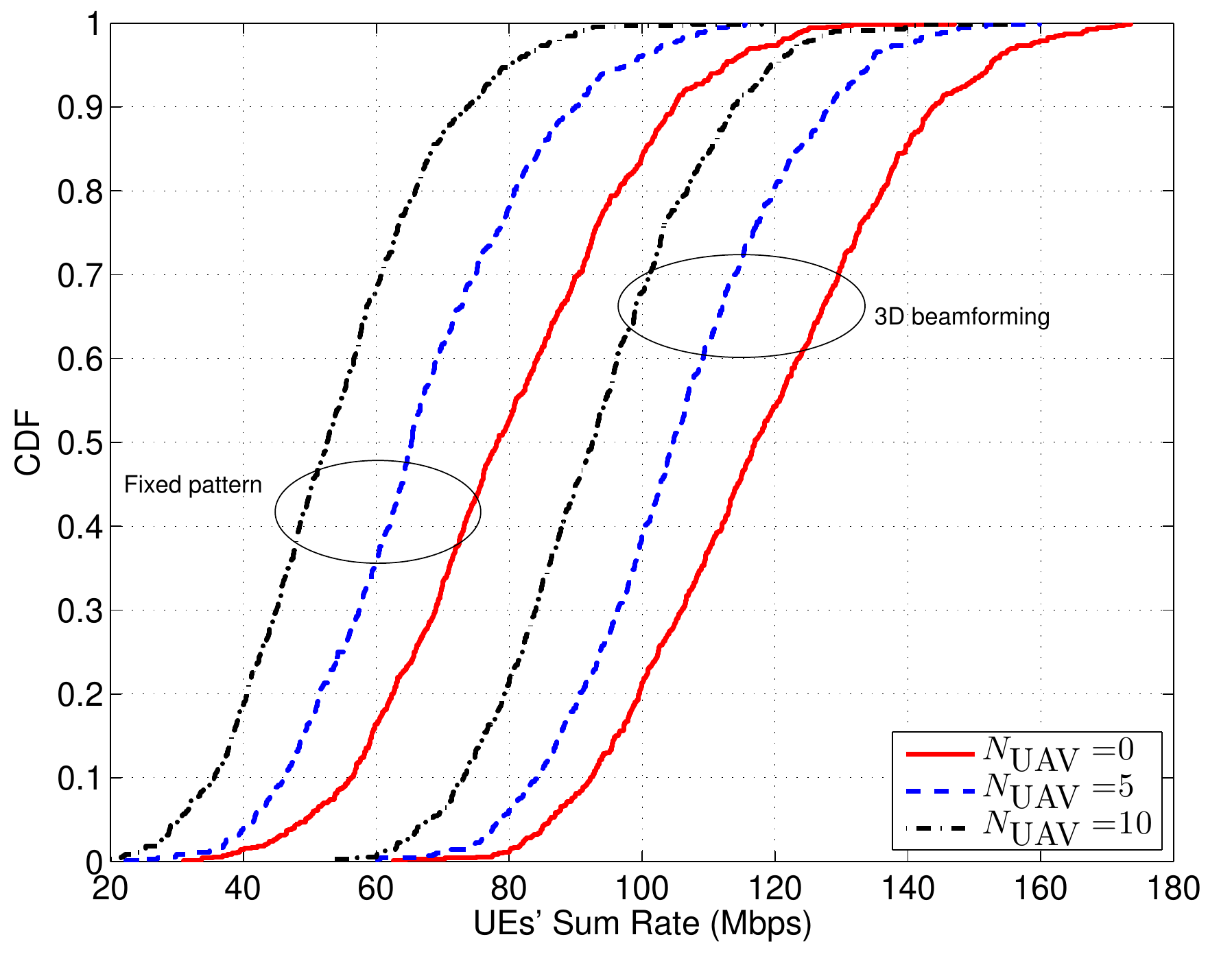}
\caption{Empirical CDF of UEs' achievable sum rate for different number of UAVs.}\label{F:SumRateCDF}
\end{figure}

Fig.~\ref{F:SumRateCDF} plots the empirical CDF of the UEs' achievable sum rate for different number of UAVs with fixed BS antenna pattern versus 3D beamforming. First, it is observed that for both array configurations, the overall system spectral efficiency degrades as the number of aerial UEs increases. This is expected since compared to ground UEs, aerial UEs suffer from more severe interference due to the higher LoS probability with the non-associated BSs, and the interference effect overwhelms the benefit of stronger direct link with its associated BS. Therefore, an aerial UE typically has poorer downlink rate performance than a ground UE, as discussed in Section~\ref{sec:newConsideration}. Thus, as the number of aerial UEs increases, the overall spectral efficiency degrades. On the other hand, Fig.~\ref{F:SumRateCDF} shows that by employing 3D beamforming even with the low-complexity MRT scheme, the system spectral efficiency can be significantly improved for all the UAV numbers. For example, for $N_{\mathrm{UAV}}=10$, the 5th percentile UEs' sum rate with fixed pattern is about 30 Mbps, while it significantly increases to more than 68 Mbps with 3D beamforming. This demonstrates the great potential of 3D beamforming for cellular systems with hybrid aerial and ground UEs \cite{978}.

\section{Conclusions and Future Directions}

In this article, we provide an overview of UAV/drone communications enabled by cellular networks, for embracing the forthcoming era of  ``Internet of Drones (IoD)''. The potential benefits of cellular-connected UAV communication as well as its unique communication and spectrum requirements are first discussed, as compared with conventional cellular communication with terrestrial users. Then, we focus on elaborating the new design considerations and promising technologies to enable future 3D heterogeneous wireless networks with both aerial and ground users. Simulation results are also provided to corroborate our discussions. Some promising directions for future research are outlined as follows.

{\bf Quality of Service-Aware Trajectory Design:}
Different from the conventional terrestrial UEs, the high and controllable mobility of UAVs offers an additional design degree of freedom via trajectory optimization for cellular-connected UAV systems. For example, for areas where ubiquitous aerial coverage by cellular BSs has not been achieved yet, the UAV trajectory should be properly  planned to avoid entering such coverage holes. 
 In general, the UAV trajectories could be jointly optimized with communication resource allocation for various performance metrics, such as spectral efficiency, or energy efficiency by taking into account the UAV's propulsion energy consumption. Another interesting direction is to develop autonomous UAVs, where the positions are self-optimized based on real-time radio measurement \cite{914}. 

{\bf Millimeter Wave Cellular-Connected UAV:}
MmWave communication that utilizes the wide available bandwidth above 28 GHz is a promising technology to achieve high-rate UAV communications \cite{948}. While mmWave communication has been extensively investigated for 5G-and-beyond cellular systems, its application in cellular-connected UAV systems faces both new opportunities and challenges. On one hand, as mmWave signals are vulnerable to blockage, the LoS-dominating UAV-BS channels offer the most favorable channel conditions for mmWave communications to be practically applied. On the other hand, the high UAV altitude and mobility requires efficient mmWave beamforming to be developed for 3D mmWave UAV-BS channels.

{\bf Cellular-Connected UAV Swarm:} UAV swarm is an effective UAV operation mode with a group of highly coordinated UAVs to complete a common mission cooperatively.  Due to the large number of UAVs and their close separations, it would be quite challenging and inefficient to connect each individual UAV directly with the cellular BSs. Instead, one promising approach is cellular-assisted U2U (UAV-to-UAV) communications, where the cellular BSs offer the backbone connectivity between the aerial network formed by the UAVs and the cellular core network. Besides, the use of massive MIMO technology for communications with UAV swarms has been recently studied in \cite{951}. More research efforts are needed to investigate the most effective aerial network topologies and the seamless integration of the aerial and cellular networks.

\bibliographystyle{IEEEtran}
\bibliography{IEEEabrv,IEEEfull}

\end{document}